# Experimentally demonstrated an unidirectional electromagnetic cloak designed by topology optimization

Lu Lan,[1)] Fei Sun,[1,2)] Yichao Liu,[1,3)] C. K. Ong [3)] and Yungui Ma[1,a)]

[1]*Centre for Optical and Electromagnetic Research, JORCEP, State Key Lab of Modern Optical Instrumentation, Zhejiang University, Hangzhou 310058, China*
[2]*Department of Electromagnetic Engineering School of Electrical Engineering, Royal Institute of Technology (KTH), S-10044 Stockholm, Sweden*
[3]*Centre for Superconducting and Magnetic Materials, Department of Physics, National University of Singapore, 2 Science Drive 3, Singapore 117542*

Electromagnetic invisible devices usually designed by transformation optics are rather complicated in material parameters and not suitable for general applications. Recently a topology optimized cloak based on level-set method was proposed to realize nearly perfect cloaking by Fujii et al [Appl. Phys. Lett. 102, 251106 (2013)]. In this work we experimentally implemented this idea and fabricated a unidirectional cloak with a relative large invisible region made of single dielectric material. Good cloaking performance was verified through measurement which consists very well with numerical simulation. The advantages and disadvantages of this optimization method are also discussed.

Aug 26, 2013

[a)] Corresponding author: yungui@zju.edu.cn

Electromagnetic (EM) invisibility, generally regarded as a major advancement of modern electromagnetism, has triggered great research interests and experienced rapid development in the past few years since the pioneering theoretical works independently done by Pendry and Leonhardt.[1,2] Transformation optics (TO) technique has been basically employed to manipulate the light path and design the invisible device.[3,4] The general TO method as initially developed by Pendry[1] and later used by the first experiment[5] and some other works[6-8] led to anisotropic inhomogeneous media with complex material parameters that are very difficult for implementation. Later quasi-conformal mapping [9] or bilinear transformation technique[10] has been developed to design weakly anisotropic cloaking media. After certain approximations and some simplifications, the TO-based devices could be realized by artificial composites or natural crystals.[11-17] However the performance of these reduced devices are far from the ideal cases. Besides, it is known that these TO-based methods will produce inhomogeneous media of complex dielectric profiles, in particular with some local superluminal parametric requirement, which make them practically very disadvantageous. On the other hand, other approaches without coordinate transformation have also been investigated in the literature to achieve a more material-sound and feasible cloak.[18-22] Among them, the most promising one seems to be the computer optimization method that relies on numerical simulation and some optimization algorithms to find a best cloaking structure.[21] Although it is less of mathematical physics, this method does provide us an efficient way to have an easily realizable cloak.

Computer optimization usually uses energy functional as a target function and looks for its optimized medium by iteratively modifying the cloaking model, i.e., varying either composition or topology to achieve required gradient index. Composition (or equally refractive index) optimization with given structures is efficient to reach the target but it may lead to some extreme or unphysical material parameters. Recently Chen et al managed to use this method to realize an anisotropic cloaking shell without superluminal limit and demonstrate good cloaking performance in microwaves.[23] On the other hand topology optimization with given materials of available parameters (i.e., permittivity $\varepsilon > 1$) is another option to find a suitable cloaking structure.[21] The resultant cloaking topology may be rather complicated but usually is not a challenge for current fabrication technology. This method is more promising to yield a practical invisible cloak for a different spectrum, especially for some special cases, e.g., where only certain direction of cloaking is required.[21]

In this work, we experimentally demonstrate an unidirectional EM cloak in microwaves designed by a topology optimization method according to the methodology theoretically developed by Fujii et al.[24] Unidirectional cloak was first proposed by Cui et al using the general TO method from a line transformation[25] and later experimentally demonstrated by the same authors but using weakly anisotropic gradient media designed by the quasi-conformal mapping.[26] Recently, by using the bilinear transformation, Landy et al demonstrated similar unidirectional cloaking effect made of anisotropic magnetic metamaterials.[27] In a very recent work, the authors utilized purely conformal mapping to design and fabricate an isotropic unidirectional cloak that offers low loss and wide band response capabilities.[28] But, all these transformed media are of complex index profiles locally with superluminal parametrical requirement that make them unsuitable for practical applications. In this letter, a relative complex cloaking topology but with a commercial available material, Teflon with dielectric constant ($\varepsilon$ = 2) and having a relatively large invisible region is fabricated. The unidirectional cloaking performance is experimentally verified in comparison with numerical simulations.

There are two different numerical schemes for topology optimization in cloaking design.[21,24] Andkjær et al firstly used the density filter to obtain a nonmagnetic cloak of gradient index profile [21] and later presented a discrete dielectric cloak using a Heaviside filter in optimization, [29] which has been followed by Urzhumov et al to make an experiment demonstration.[30] In this work we utilized a dielectric cloak design by the level-set based topology optimization which was first theoretically proposed by Fujii et al,[24] which allows us to have a clear cloak configuration more directly with a single dielectric material, thus greatly facilitating the fabrication. Figure 1 gives a schematic of the design model adapted from Ref 24. The light grey part of an outer radius $R_2$(= 4.5$\lambda$, $\lambda$ is the free-space wavelength) denotes the design domain ($\Omega_d$) including the cloaking medium (blue color). The dark grey disk of radius $R_1$(= 1.5$\lambda$) is the invisible region represented by a perfectly electric conducting (PEC) metal cylinder. An outmost perfectly matched layer (PML) is used to mimic an infinitely large air space enclosing the design domain. The objective function($\psi_s$) is defined by the intensity integral of the scattered electric field ($E_s$) along the dashed brown line under a plane-wave excitation ($E_i$), i.e., $\psi_s = \frac{1}{\psi_0} \int_{\partial\Omega} E_s E_s^* d\Omega$, where $\psi_0$ is the scattered field intensity when there is no cloak around the metal and $E_s^*$ is the complex conjugate of $E_s$. The dielectric profile of the design domain is represented by $\varepsilon(x) = \varepsilon_{air} + \chi \cdot \varepsilon_d$ where $\chi$ = 1 or 0 dependent on the range of a level-set function defined as piecewise constant values to dielectric material boundaries. A finite element method (FEM) is used to calculate the scattered field and update the level-set functions. For the technical details about this optimization method the readers may refer to the theoretical papers published before.[24] Our adapted optimized unidirectional cloaking system, as shown in Fig. 1, looks like an 'eye' structure with the blue 'eyelid' as the cloak and the dark 'eyeball' as the invisible region. In the following simulation and experiment, we take the working wavelength $\lambda$ = 20 mm (i.e., 15 GHz) and $\varepsilon_d$ = 2, which is very suitable for implementation.

Figure 2(a) shows one snapshot of the simulated wave pattern of the optimized cloak with the incident wave coming from the left side. It's seen that the electromagnetic wave is guided and delayed by the constant dielectric cloak and exits in phase with the background wave that does not interact with the cloak. The straight wavefronts around the cloak are slightly disturbed indicating an imperfect cloaking, which can be further improved by tuning the optimization factor.[24] In this design impedance mismatch at the dielectric-air boundary will not be an issue but the total phase accumulation will influence the final cloaking performance. Figure 2(b) plots the biscattering curve of the simulated wave profile, i.e., the angle-dependent far-field radiation pattern for electric field normalized by the incident value. Prominent cloaking performance is evidenced by the figure. Small amplitude of irregular side scattering is also observed.

In experiment we used the low loss dielectric material of Teflon ($\varepsilon = 2$ around 15 GHz) as the basic ingredient and fabricated such a cloaking sample by a precise engraving machine (Jingdiao CNC). The fabrication took about 15 min to complete and is rather efficient especially compared to the implementation for the complicated TO devices. Figure 3 gives a top view of the fabricated sample mainly consisting of two dielectric strips and one central aluminum disk. Such a cloaking setup won't be a big problem to replicate in the THz or even optical spectrum. To measure the sample we put it inside a home-made parallel-plate two-dimensional field mapping platform which is driven by a program controlled *xy*-step motor.[31] The planar incident wave beam is produced through a half hyperbolic lens (aperture size = 220 mm) that fed by a line-current source at the focal point.[32] Microwave absorbing foams are placed at the edges of the platform to reduce the possible edge scattering. A 0.05 mm thick transparent plastic paper ($\varepsilon \approx 3$) is covered on the surface of the sample to inhibit the possible displacement in the moving measurement. The microwave signal is generated and processed by a vector network analyzer (RS ZVA40). The resolution of the mapping field pattern is fixed at 1 mm by controlling the step length of the driving motor.

The measured electric field patterns for a Al disk without and with a cloak are given in Figs. 4(a, b) and Figs. 4(c, d), respectively. The left/right two figures draw the real/modulus profiles of the electric field. It's clear that a bare metal disk incurs strong scattering by forming standing wave and shadow right before and after the metal as shown in Figs. 4(a) and 4(b). These disturbances are greatly suppressed in the case with a cloak as shown in Figs. 4(c) and 4(d). The straight and continuous wavefronts are basically restored after the cloak. The measured field pattern in Fig. 4(c) is in good agreement with the numerical prediction in Fig. 3(a). The slight difference might be caused by the manufacturing problem that cannot reproduce some narrow gaps exactly due to the size limit of the smallest engraving head we have. The current cloaking performance is reasonably good and acceptable compared with the previously published data for the TO devices.[26-28]

Optimization method is usually processed at a single point frequency with one polarization and one incident angle. We checked the scattering pattern at different frequencies and found relative good cloaking effect was obtained only within a narrow frequency band from 14.7 to 15.6 GHz. Our current setup is primarily designed for transverse electric (TE) polarization. The simulation shows (not given here) that the same cloak will induce an obvious shadow after the cloak for transverse magnetic (TM) polarization although it does help to reduce the metal scattering. The difference could be understood because the boundary scattering at the edges of the dielectrics is polarization dependent, which changes the total phase delay for waves passing over the cloak. Polarization insensitive topology optimized cloak has been investigated recently by simultaneously taking both $E_z$ and $H_z$ components into account.[29]

In conclusion, we have experimentally demonstrated a unidirectional cloak with a relative large invisible region made of a constant dielectric designed by topology optimization. The fabrication process of a sample is substantially simplified and also efficiently improved (it only needs 15min to fabricate the sample) compared with the optically transformed media, which offers cloaking and related technologies more promising for practical applications. In future work, pulse excitation in the optimization can be considered to find an electromagnetic cloak with broad frequency response. Multiple angles incidence can be also discussed by the same method with increased topology symmetries.

**Acknowledgments**

The authors are grateful to the partial supports from NSFCs 61271085, 60990322 and 91130004, the National High Technology Research and Development Program (863 Program) of China (No. 2012AA030402), NSF of Zhejiang Province (LY12F05005), the Program of Zhejiang Leading Team of Science and Technology Innovation, NCET, MOE SRFDP of China, the DIRP grant of Singapore (R144000304232) and AOARD.

**References**

1. J. B. Pendry, D. Schurig, and D. R. Smith, Science **312**, 1780 (2006).
2. U. Leonhardt, Optical conformal mapping, Science **312**, 1777 (2006).
3. U. Leonhardt and T. G. Philbin, Prog. Opt. **53**, 69 (2009).
4. H. Y. Chen, C. T. Chan, and P. Sheng, Nature Mater. 9, 387 (2010).
5. D. Schurig, J. J. Mock, B. J. Justice, S. A. Cummer, J. B. Pendry, A. F. Starr, and D. R. Smith, Science **324**, 977 (2006).
6. Y. G. Ma, C. K. Ong, T. Tyc, and U. Leonhardt, Nature Mater. **8**, 639 (2009).
7. H. Y. Chen and C. T. Chan, Appl. Phys. Lett. **90**, 241105 (2007).
8. Y. Lai, J. Ng, H. Y. Chen, Z. Q. Zhang, and C. T. Chan, Phys. Rev. Lett. **102**, 253902 (2009).
9. J. S. Li and J. B. Pendry, Phys. Rev. Lett. **101**, 203901 (2008).
10. Y. Luo, J. J. Zhang, H. Chen, L. X. Ruan, B-I Wu, and J. A. Kong, IEEE Trans. Antennas. Propag. **57**, 3926 (2009).
11. R. Liu, C. Ji, J. J. Mock, J. Y. Chin, T. J. Cui, and D. R. Smith, Science **323**, 366 (2009).
12. J. Valentine, J. S. Li, T. Zentgraf, G. Bartal, and X. Zhang, Nature Mater. **8**, 568 (2009).
13. L. H. Gabrielli, J. Cardenas, C. B. Poitras, and M. Lipson, Nature Photo. **3**, 461 (2009).
14. F. Zhou, Y. J. Bao, W. Cao, T. S. Colin, J. Q. Gu, W. L. Zhang, and C. Sun, Sci. Rep., 2011; DOI: 10.1038/srep00078.
15. T. Ergin, N. Stenger, P. Brenner, L. B. Pendry, and M. Wegener, Science **328**, 337 (2010).
16. H. F. Ma and T. J. Cui, Nature Comm. **1**, 124 (2011).
17. N. Wang, Y. G. Ma, R. F. Huang, and C. K. Ong, Opt. Expr. **21**, 5941 (2013).
18. G. Milton and N. A. Nicorovici, Proc. Royal Soc. A **462**, 3027 (2006).
19. A. Alù and N. Engheta, Phys. Rev. Lett. **100**, 113901 (2008).
20. F. G. Vasquez, G. W. Milton, and D. Onofrei, Phys. Rev. Lett. **103**, 073901 (2009).
21. J. Andkajaer and O. Sigmund, Appl. Phys. Lett. **98**, 021112 (2011).
22. L. Sanchis, V. M. Garcı́a-Chocano, R. Llopis-Pontiveros, A. Climente, J. Martı́nez-Pastor, F. Cervera, and J. Sánchez-Dehesa, Phys. Rev. Lett. **110**, 124301 (2013).
23. S. Xu, X. Cheng, S. Xi, R. Zhang, H. O. Moser, Z. Shen, Y. Xu, Z. Huang, X. Zhang, F. Yu, B. L. Zhang, and H. Chen, Phys. Rev. Lett. **109**, 223903 (2012).
24. G. Fujii, H. Watanabe, T. Yamada, T. Ueta, and M. Mizuno, Appl. Phys. Lett. **102**, 251106 (2013).
25. W. X. Jiang, H. F. Ma, Q. Chen, and T. J. Cui, J. Appl. Phys. **107**, 034911 (2010).
26. H. F. Ma, W. X. Jiang, X. M. Yang, X. Y. Zhou, and T. J. Cui, Opt. Expr. **12**, 19947 (2009).
27. Landy N. & Smith D. R., A full-parameter unidirectional metamaterial cloak for microwaves, Nature Mater. **12**, 25 (2013).
28. Y. G. Ma, Y. C. Liu, L. Lan, T. T. Wu, W. Jiang, C. K. Ong, and S. L. He, Sci.

Rep. **3**: 2182 | DOI: 10.1038/srep02182 (2013).

[29] J. Andkjær, N. A. Mortensen, and O. Sigmund, Appl. Phys. Lett. **100**, 101106 (2012).

[30] Y. Urzhumov, N. Landy, T. Driscoll, D. Basov, and D. R. Smith, Opt. Lett. **38**, 1606 (2013).

[31] L. Zhao, X. Chen, and C.K. Ong, Rev. Sci. Instr. **79**, 124701 (2008).

[32] B. J. Justice, J. J. Mock, L. Guo, A. Degiron, D. Schurig, and D. R. Smith, Opt. Expr. **14**, 8694 (2006).

**Captions**

Figure 1 (Color online) Schematic of the optimization model adapted from ref (24). The light grey region of outer radius $R_2$ describes the design domain that encloses the central PEC cylinder representing the invisible region (of radius $R_1$). The dashed brown line within the outer air is used to integrate the intensity of the scattered field ($E_s$) defined as the objective function under the plane wave excitation ($E_i$) from the left side. The discontinuous structure highlighted in the blue color is the optimized cloak designed here under the condition $R_1= 1.5\lambda$, $R_2= 4.5\lambda$ and $\varepsilon_d= 2$.

Figure 2 (Color online) Simulation of the scattered wave pattern. (a) One snapshot of the electric field pattern and (b) the corresponding biscattering curve. The simulation frequency is 15 GHz. A rainbow color scheme is employed to represent the field values/signs and the wave feature. The biscattering curve is calculated from the far-field radiation field pattern which is derived from the field along the dashed brown line shown in Fig. 1.

Figure 3 (Color online) Device photo. The white 'eyelid' shaped cloak is made of Teflon which has permittivity about 2 near 15 GHz. The center invisible region is replaced by an aluminum disk of diameter = 60 mm =$3\lambda$. The sample is 5 mm in thickness. In measurement there is no air gap between the sample surface and the measurement waveguide plates.

Figure 4 (Color online) Measured results. (a) and (b) are the real and magnitude patterns of a reference case for a metal disk without cloak, respectively. (c) and (d) are the real and magnitude patterns of the case with cloak, respectively. The electric field is characterized here. A hyperbolic dielectric lens is used to transfer a line current source into the planar incident wave at 15 GHz.

Figure 1

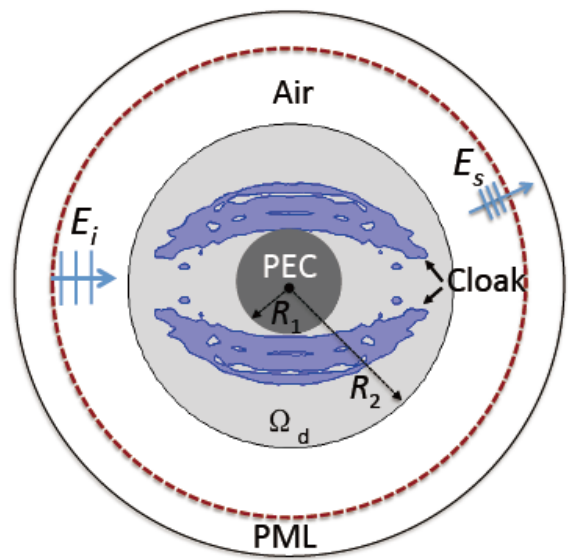

Figure 2

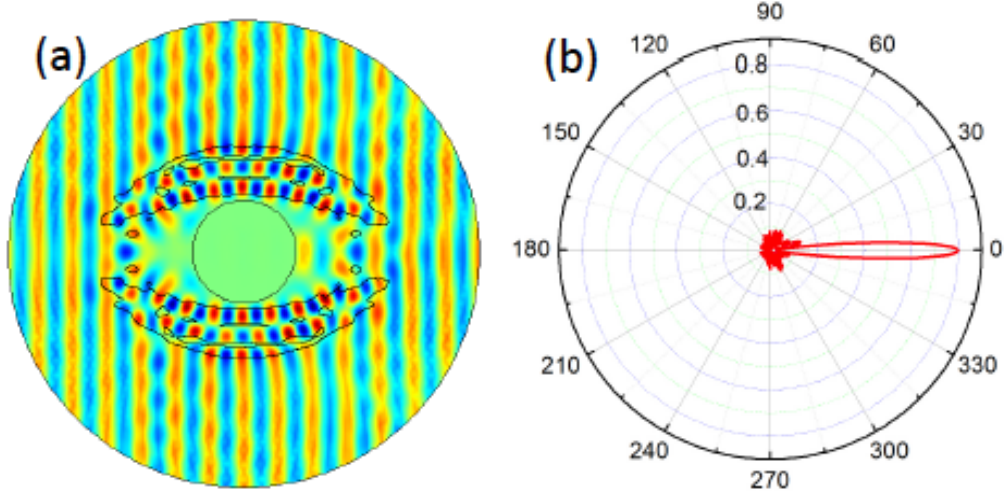

Figure 3

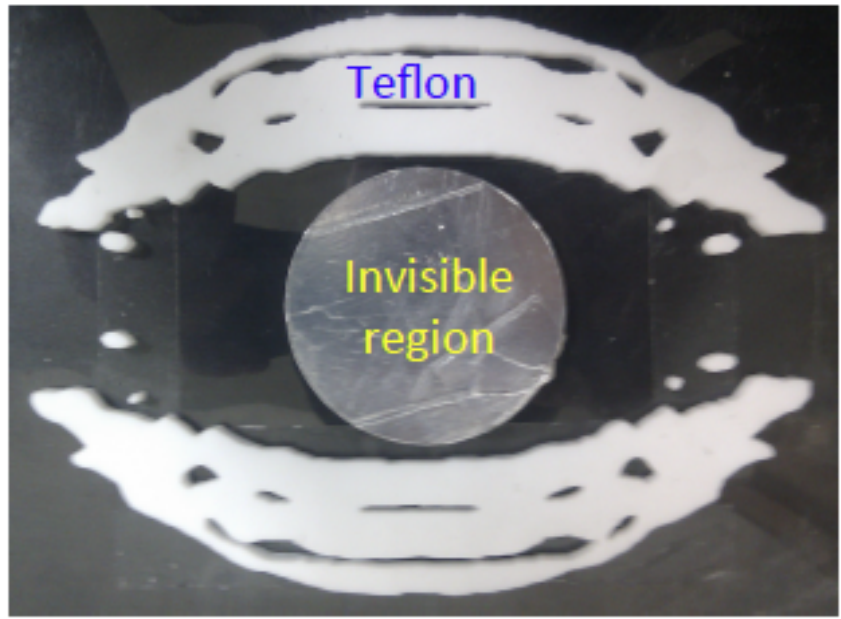

Figure 4

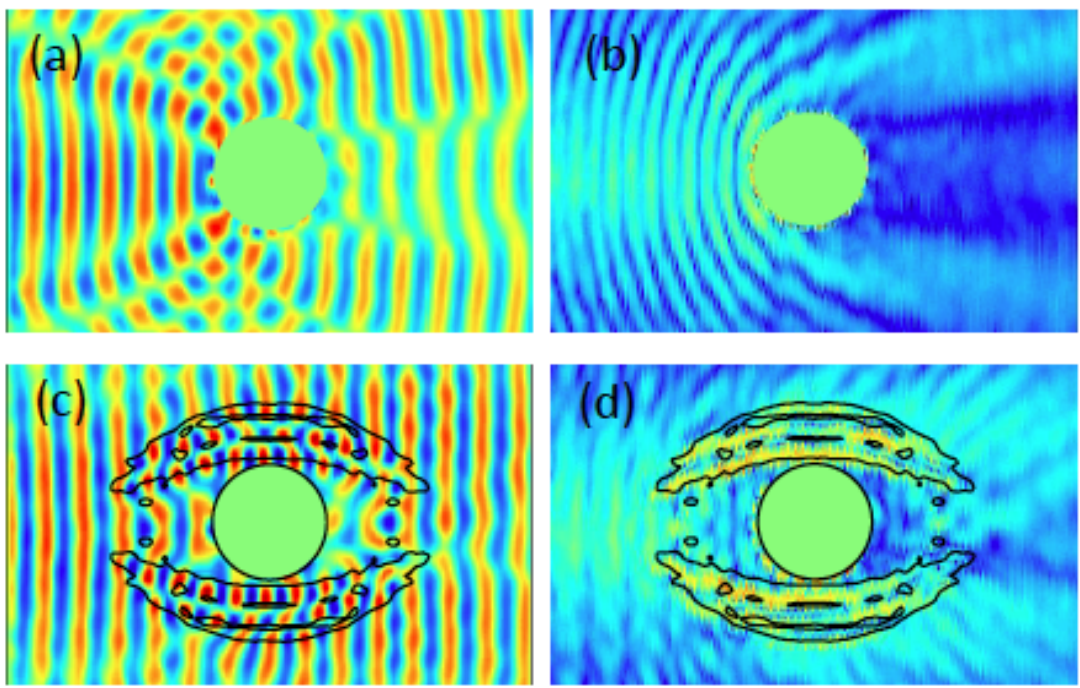